\documentstyle[preprint,prb,aps,epsfig]{revtex}
\begin{document}
\draft
\title{Intrinsic noise of a micro-mechanical displacement detector
based on the radio-frequency single-electron transistor}

\author{Yong Zhang and Miles P. Blencowe}
\address{Department of Physics and Astronomy, Dartmouth College, 
Hanover, New Hampshire 03755}
\date{\today}
\maketitle
\begin{abstract}
We investigate the intrinsic noise of a micro-mechanical 
displacement detector based on the radio-frequency single-electron 
transistor (rf-SET). Using the noise analysis of a SET by Korotkov as our 
starting point, we determine the spectral density of the displacement noise 
due to the tunneling current shot noise. The resulting mechanical displacement 
noise decreases in inverse proportion to the increasing gate voltage. 
In contrast, the displacement noise due to the fluctuating SET island charge 
increases approximately linearly with increasing gate voltage. Taking into 
account both of these noise sources results in an optimum gate voltage value 
for the lowest displacement noise and hence best sensitivity. We show that a 
displacement sensitivity  of about $10^{-4}~{\rm \AA}$ and a force sensitivity  
of about $10^{-16}~{\rm N}$ are 
predicted for a micron-sized cantilever with a realizable resonant frequency 
 $100~{\rm MHz}$ and quality factor $Q\sim~10^{4}$. Such sensitivities 
would allow the 
detection of quantum squeezing in the mechanical motion of the micro-mechanical 
cantilever and the detection of single-spin magnetic resonance in magnetic 
resonance force microscopy (MRFM).
\end{abstract}


\section{Introduction}

Fast and ultra-sensitive displacement and force detection is of 
great interest for a broad range of fundamental applications. It is an essential 
requirement for the proposed detection of quantum 
superposition states\cite{bose}, quantum 
squeezed states\cite{blencowe}, and single-spin 
magnetic resonance in 
MRFM\cite{sidles} which has also been proposed for the read-out stage of a 
solid-state NMR quantum computer.\cite{berman} The common 
requirement of these examples is the need to perform the measurement using a 
micro-to-nanoscale-mechanical detector at, or even below, 
the ``standard quantum limit" 
(SQL) $\Delta x_{\rm SQL}=(\hbar/2m\omega_{0})^{1/2}$,
and in the 
radio-frequency range. Considering, for instance, a micron-sized cantilever 
with a resonant frequency of $100~{\rm MHz}$ and quality factor $Q\sim~10^{4}$, 
the detection of 
quantum squeezed states requires a displacement sensitivity of about 
$10^{-4}~{\rm \AA}$.\cite{blencowe} As another example, 
in detecting single-spin 
magnetic resonance, the interaction force between a nano-magnet with a 
targeted electron spin is typically on the order of about 
$10^{-16}~{\rm N} (100~{\rm aN})$.\cite{sidles} 
If a micron-sized cantilever on which the nano-magnet is mounted is directly 
coupled at the spin precession frequency ($100~{\rm MHz}$ or higher), 
the magnetic force will cause the cantilever to oscillate at an 
amplitude of about $10^{-4}~{\rm\AA}$ for a single spin.

Following the development of the 
rf-SET,\cite{schoelkopf,fujisawa,aassime} 
it has been proposed 
to use an rf-SET-based displacement detector for measuring sub-angstrom motion of 
micron-scale and smaller mechanical oscillators.\cite{blencowe2} In the rf-SET, 
 $1/f$ noise which limits the applications of the conventional 
dc-SET is considerably improved. A shot noise limit of 
$10^{-6}~{\rm\AA}/\sqrt{Hz}$ may be 
achievable by using an rf-SET displacement detector,\cite{blencowe2} 
provided the coupling 
(gate voltage) between the SET and micro-mechanical oscillator is strong 
enough.
However,  the back-action of the SET onto the mechanical 
oscillator must also be taken into account:\cite{schwab} 
the fluctuating force acting on the oscillator 
due to the fluctuating island charge produces another contribution to the 
displacement noise. The total displacement noise will be the sum of 
contributions from two sources - the shot noise in the tunneling current and 
the force noise due to the fluctuating island charge.

In this paper, we give a  calculation of the displacement noise 
of the rf-SET displacement detector. By using the intrinsic noise analysis 
method of Korotkov,\cite{korotkov} the spectral density of the shot 
noise $S_{I}$ in the tunneling 
current and the spectral density of the force noise $S_{F}$ are determined, and 
the 
total spectral density of the displacement noise $S_{X}$ is then deduced. We also 
present two simple expressions from which  good estimates  of the  displacement 
noise due to the shot noise and the fluctuating 
island charge can be obtained. Results for the ultimate intrinsic 
noise limit of the detector for different parameter values are presented and 
discussed. The displacement and force sensitivities are estimated for an 
optimized rf-SET displacement detector and we find that they meet the 
requirement for detecting quantum squeezing and single-spin magnetic resonance.

\section{Intrinsic noise of the rf-SET displacement detector}

The scheme of the rf-SET displacement detector is shown in Fig.~1.
The basic principle of the 
device involves locating one of the  gate 
capacitor plates of the SET  on the cantilever so that, 
for  fixed gate voltage bias,  a mechanical 
displacement is converted into a polarization charge fluctuation. 
The  
stray capacitance $C_{s}$ of the leads contacting the SET and an inductor 
 $L$ form a  tank circuit with resonant  frequency 
$\omega_{T}=(LC_{s})^{-1/2}$, loaded by the SET. A 
monochromatic carrier wave is sent down the cable. At 
the resonant frequency the circuit impedance is small and the 
reflected power provides a measure of the SET's differential 
resistance $R_{d}$.\cite{korotkov2} When the gate capacitor is biased,
mechanical motion of the cantilever is converted into differential 
resistance changes, hence modulating the reflected signal power.

To simplify the algebra and make the final results concise, 
 we shall in fact consider 
a symmetric dc-SET in the calculations  with junction resistances 
$R_{1}= R_{2}= R_{j}$ and 
junction capacitances $C_{1}=C_{2}= C_{j}$. The differences between the 
rf- and dc-SET noise formulae are inessential, with appropriate 
time-averages required in the former\cite{blencowe2}
resulting in only small quantitative differences in the noise values.

Referring to Fig.~1, when the drain-source voltage $V_{\rm ds}$ is small 
compared to 
the voltage $e/C_{\Sigma}$, where $C_{\Sigma}=2 C_{j}+C_{g}$ is the 
total capacitance of the SET, and also the thermal energy $k_{B}T\ll 
eV_{\rm ds}$, the peaks of the tunneling 
current $I_{\rm ds}$ will be well separated in gate voltage $V_{g}$. To 
derive an expression for $I_{\rm ds}$  in the tunneling region, we 
consider only the two most probable island electron numbers $n$ and 
$n+1$,  resulting in the following approximation:\cite{blencowe2}
\begin{equation}
    I_{\rm ds}= 
    e[b_{1}(n)-t_{1}(n)]\sigma(n)+e[b_{1}(n+1)-t_{1}(n+1)]\sigma(n+1).
    \label{current1}
\end{equation}
The probabilities $\sigma(n)$ and $\sigma(n+1)$ are also given there and the 
tunneling 
rates $b_{i}(t_{i})$ from the bottom (top) across the $i$th junction 
of the rf-SET take 
the usual form [see expression (\ref{apprates}) in the appendix].

In the previous calculation of the sensitivity of the rf-SET based detector, 
the shot noise formula $S_{I} = 2eI_{\rm ds}$ was 
used.\cite{blencowe2,korotkov2} This expression, 
which gives the maximum current noise, is only approximately valid 
when $V_{\rm ds}$ is close to the tunneling threshold. 
Taking as our tool the 
intrinsic noise analysis of the SET by Korotkov, we obtain the following 
more accurate expression for the spectral density of the current noise 
$S_{I}(\omega)$ [see  appendix A for an outline of the deviation]:
\begin{equation}
S_{I}(\omega)=\frac{e^{2}a(n)b(n+1)}{a(n)+b(n+1)}+\frac{e^{2}[f(n)-g(n+1)]}
{a(n)+b(n+1)}\frac{a^{2}(n) g(n+1) -b^{2}(n+1) 
f(n)}{[a(n)+b(n+1)]^{2}+\omega^{2}},
\label{currentnoise}
\end{equation}
where
\begin{eqnarray}
    a(m)&=&b_{1}(m)+t_{2}(m),\cr
    b(m)&=&b_{2}(m)+t_{1}(m),\cr
    f(m)&=&b_{1}(m)-t_{2}(m),\cr
    g(m)&=&b_{2}(m)-t_{1}(m).
    \label{def1}
\end{eqnarray}

The spectral density of the displacement noise due to the shot noise in 
the tunneling current is:
\begin{equation}
    S_{X}^{I}(\omega)=S_{I}(\omega)/(dI_{\rm ds}/dx)^{2}.
    \label{SXI}
\end{equation}
The dependence on the displacement $x$ enters through the gate 
capacitance which is approximately $C_{g}\approx C_{g}^{0}(1-x/d)$ 
for $|x|\ll d$ where $d$ is the cantilever electrode-island electrode 
gap.

The fluctuation of the island charge will produce a 
fluctuating force on 
the cantilever, therefore leading to another contribution to the 
displacement noise. The ultimate force noise is only determined by 
the stochastic character of the tunneling process. By using the 
same method of intrinsic noise analysis,\cite{korotkov}
the following expressions are 
obtained for the spectral density of the force noise $S_{F}(\omega)$ 
[see  appendix B for an outline of the deviation]:
\begin{eqnarray}
    S_{F}(\omega)=&&\frac{4A^{2}e^{4}a(n)b(n+1)}{a(n)+b(n+1)}\cr
    &&\times \frac{(2C_{j}/C_{g})^{2} (2n+1+2\Delta n)^{2} 
    +(4C_{j}/C_{g})(2n+1)(2n+1+2\Delta 
    n)+(2n+1)^{2}}{[a(n)+b(n+1)]^{2}+\omega^{2}},
    \label{SF}
\end{eqnarray}
where \begin{equation}
A=C_{g}(2C_{j}-C_{g})/2dC_{\Sigma}^{3}
\end{equation}
and
\begin{equation}
    \Delta n=\frac{C_{g}V_{g}}{e}-\frac{C_{g}V_{\rm ds}}{2e}-n-\frac{1}{2}.
\end{equation}

We model the cantilever as a simple harmonic oscillator, so that the 
spectral density of the displacement noise due to the fluctuating 
force is:
\begin{equation}
    S_{X}^{F}(\omega)=\frac{S_{F}(\omega)/m_{\rm 
    eff}^{2}}{(\omega^{2}-\omega_{0}^{2})^{2}+
    \omega^{2}\omega_{0}^{2}/Q^{2}},
    \label{SXF}
\end{equation}
where $m_{\rm eff}$ is the motional mass of the cantilever, 
$\omega_{0}$ is its resonant frequency, and $Q$ is 
the quality factor. For a cantilever with geometry $l\times w\times 
t$ (length$\times$width$\times$thickness)  and made from material of 
density $\rho$ and Young's modulus $E$, the resonant frequency of the 
lowest-order flexural mode is\cite{sidles}
\begin{equation}
    \omega_{0}=3.516 \frac{t}{l^{2}} \sqrt{\frac{E}{12\rho}}.
    \label{frequency}
\end{equation}
The motional mass of the cantilever is given by $m_{\rm eff}=\rho 
lwt/4$ and the effective spring constant $k_{\rm eff}$ at the 
cantilever's tip is $k_{\rm eff}=m_{\rm eff}\omega_{0}^{2}$.
The total displacement noise will be given as the sum of the above 
two contributions (\ref{SXI}) and (\ref{SXF}):
\begin{equation}
    S_{X}(\omega)=S_{X}^{I}(\omega)+S_{X}^{F}(\omega).
    \label{SXtot}
\end{equation}

In addition, the following expressions are given here only for simple 
estimates of the displacement noise at the resonant frequency of the 
cantilever [see  appendix C for an outline of their derivation]:
\begin{equation}
    S_{X}^{I}=K\frac{2R_{j}}{e V_{\rm ds}}\left[\frac{2.5 
    k_{B}TdC_{\Sigma}}{C_{g}(V_{g}-V_{\rm ds}/2)}\right]^{2},
    \label{approxSXI}
\end{equation}
\begin{equation}
    S_{X}^{F}=\frac{Q^2 e^{3}R_{j}}{k_{\rm eff}^{2}V_{\rm 
    ds}}\left[\frac{C_{g}(2C_{j}-C_{g})(V_{g}-V_{\rm ds}/2)}
    {d C_{\Sigma}^{2}}\right]^{2},
    \label{approxSXF}
\end{equation}
where $K\approx 40$ is a constant.

\section{Results and Discussion}

In this paper, we consider crystalline Si cantilevers with mass 
density $\rho= 2.33\times 10^{3}~{\rm kg/m}^{3}$ and Young's modulus 
$E_{100} = 1.33\times 10^{11}~{\rm N/m}^{2}$. 
To make the picture clear, we compare the displacement noise for 
different cantilevers. Table I lists the resonant frequencies $f_{0}$ and 
masses $m$ (the effective mass of a cantilever is $m_{\rm eff}= m/4$) of 
cantilevers 
with the same width and thickness but different lengths. The resonant 
frequencies $f_{0}= \omega_{0}/2\pi$ are obtained from expression 
(\ref{frequency}). We 
mainly focus our attention on the detector using the cantilever 
with geometry $1.25\times 0.6\times 0.15$ in $\mu{\rm m}$, quality factor 
$Q=10^{4}$, resonant 
frequency $f_{0}= 117~{\rm MHz}$, and effective spring constant 
$k_{\rm eff} = 3.6\times 10^{9}~{\rm aN/\AA}$. The 
standard quantum limit of the fundamental flexural mode of this cantilever is 
$\Delta x_{\rm SQL}= 3.3\times 10^{-4}~{\rm \AA}$, so that the minimum detectable
force at the resonant frequency is $\Delta F= k_{\rm eff}\Delta x_{\rm 
SQL}/Q= 120~{\rm aN}$.

Fig.~2 shows the dependence of the displacement noise $\sqrt{S_{X}^{I}}$ on 
the gate voltage $V_{g}$ at the resonant 
frequency $f_{0}= 117~{\rm MHz}$ of cantilever . The result 
assumes a symmetric SET with junction capacitance $C_{j}= 100~{\rm aF}$, 
junction resistance $R_{j}= 50~{\rm k}\Omega$, gate capacitance 
$C_{g}= 50~{\rm aF}$, gate
capacitor gap $d= 0.1~\mu{\rm m}$, source-drain voltage 
$V_{\rm ds}= 0.38~{\rm mV} (0.6e/C_{\Sigma})$, and 
temperature $T= 74~{\rm mK}(0.01e^{2}/C_{\Sigma})$, which are all 
routinely achievable. 
The solid line is obtained by using expressions 
(\ref{current1}), (\ref{currentnoise}), and (\ref{SXI}) and the dashed 
line using (\ref{approxSXI}). The minimum displacement noise occurs near the 
edges of the tunneling current $I_{\rm ds}$, and decreases approximately in 
inverse proportion to increasing gate voltage. The displacement noise diverges 
in the center of tunneling current peaks and in the Coulomb blockade region where the 
tunneling current is suppressed. The result here is similar to  Fig.~2 
in Ref.\ \onlinecite{blencowe2}.

The dependence of the displacement noise $\sqrt{S_{X}^{F}}$ on the gate voltage 
$V_{g}$ at the resonant frequency $f_{0}= 117~{\rm MHz}$ is illustrated in 
Fig.~3. The result 
assumes the same parameters as above. The solid line is obtained by 
using expressions (\ref{SF}) and (\ref{SXF}), and the dashed line using 
(\ref{approxSXF}). The maximum displacement noise increases linearly with 
the increasing gate voltage. 

Fig.~4 shows the total displacement noise 
$\sqrt{S_{X}}$ as a function of gate voltage. 
The tunneling current is shown as the dashed lines.
Considering the opposite dependences of $S_{X}^{I}$ and $S_{X}^{F}$ on 
gate voltage, we will expect an optimum gate voltage value for the 
lowest displacement noise and best sensitivity. 
But there are two problems involved in the determination. 
The first problem is that near the edges the tunneling current $I_{\rm 
ds}$ may be too small to be measured accurately. 
The second  is that because $S_{X}^{I}$ 
and $S_{X}^{F}$ change in different ways with  increasing gate 
voltage $V_{g}$, it is rather involved finding the most optimum value of 
$V_{g}$ at which the total 
noise is minimum; the positions of the minima in the displacement noise 
 doublets will change relatively to the tunneling current peaks with 
increasing gate voltage. An alternative, simpler strategy is to investigate the 
displacement noise at the gate voltage for which the tunneling current is  
fixed and measurable, say between $30\%$ and $60\%$ 
of its maximum value $I_{\rm ds, max}$. 
Fig.~5 shows the total displacement noise $\sqrt{S_{X}}$ for our rf-SET 
displacement detector using different cantilever parameter values 
listed in Table I. The above method 
is utilized with the displacement noise determined where the tunneling 
current is half its maximum value. For the detector using a cantilever 
with resonant frequency $f_{0}= 117~{\rm MHz}$, the minimum displacement noise 
is $4.3\times 10^{-6}~{\rm \AA}/\sqrt{Hz}$, corresponding to an absolute 
displacement sensitivity of $5\times 10^{-4}~{\rm\AA}$ and force sensitivity 
$180~{\rm aN}$. Note that the minimum displacement 
noise increases and shifts to  lower gate voltage for increasing 
cantilever length. As is evident from (\ref{approxSXF}), this is a direct 
consequence of the effective spring constant dependence: 
obviously a stiffer cantilever will be displaced less by the same applied force.

Fig.~6 shows the detector displacement noise for  different 
junction capacitance values. Both $S_{X}^{I}$ and $S_{X}^{F}$ 
decrease with decreasing $C_{j}$. 
Because small displacements are 
detected by monitoring changes in tunneling current, taking a smaller 
junction capacitance  serves another advantage 
for the experiment: with $C_{\Sigma}V_{\rm ds}/e$ 
kept fixed, a smaller junction capacitance corresponds to a larger 
source-drain voltage, so that the maximum tunneling current $I_{\rm ds, max}$ increases 
according to expression (\ref{currentmax}).  

Fig.~7 
shows the displacement noise when the tunneling current is biased at 
various values between 
$30{\%}$ and $60{\%}$ of $I_{\rm ds, max}$. 
We see that performing detection at a smaller percentage of the maximum 
tunneling current becomes another effective way to further decrease the 
displacement noise. The lowest displacement noise we 
obtain is $\sqrt{S_{X}} = 3.2\times 10^{-6}~{\rm\AA}/\sqrt{{\rm Hz}}$, 
assuming $C_{j}= 100~{\rm aF}$, $C_{g}= 50~{\rm aF}$, and $I_{\rm ds}= 
0.57~{\rm nA}$, biased at $30{\%}$ of its maximum value. 
The optimum gate voltage is $1.7~{\rm V}$ in 
this case. 

In conclusion, approaching the SQL will become possible for a carefully 
engineered rf-SET displacement detector. A sensitivity at, or even 
beyond the 
SQL may be achievable by using the same parameters as in the above 
analysis but with 
the smaller junction capacitance $C_{j} = 75~{\rm aF}$. 
Such a sensitivity would enable us to observe quantum squeezed states in 
micro-mechanical systems. The proposed 
displacement detector also meets the crucial requirement for advances in MRFM. 
The radio-frequency micro-mechanical cantilever used in our model allows 
direct coupling at the spin precession frequency, therefore allowing fast 
read-out, while the predicted force sensitivity would allow the measurement of 
single-spin resonance. 

\acknowledgements

The authors thank Keith Schwab and Rob Schoelkopf for pointing out the 
need to include the backaction on the cantilever due to the fluctuating island 
charge and Martin Wybourne for helpful 
discussions. This work was supported in part  by the National Security 
Agency (NSA) and Advanced Research and Development Activity (ARDA) under 
Army Research Office (ARO) contract number DAAG190110696, and by an award from 
Research Corporation. 

\appendix
\section{Tunneling current shot noise}

The tunneling current shot noise is given by expression (30) in Ref.\ 
\onlinecite{korotkov}. In our notation, it is:
\begin{eqnarray}
    S_{I}(\omega)=&&2U+e^{2} 
    \sum_{m,m'}\left[t_{2}(m')-b_{2}(m')+t_{1}(m')-b_{1}(m')\right] 
    B_{m',m}\cr
    &&\times\left\{\left[t_{2}(m-1)-b_{1}(m-1)\right]\sigma(m-1)+
    \left[t_{1}(m+1)-b_{2}(m+1)\right]\sigma(m+1)\right\},
    \label{appSI}
\end{eqnarray}
where
\begin{equation}
2U=(e^{2}/2)\sum_{m}\sigma(m)\left[t_{2}(m)+b_{2}(m)+t_{1}(m)+b_{1}(m)\right]
\label{U}
\end{equation}
and $B_{m',m}$ is the real part of the inverse of the tridiagonal 
matrix $(i\omega I -\Gamma)$, where $I$ is the unit matrix  and
\begin{equation}
\Gamma_{nm}=\delta_{n-1,m}a(m)+\delta_{n+1,m}b(m)-\delta_{n,m}c(m),
\label{Gamma}
\end{equation}
with
\begin{eqnarray}
    a(m)&=&b_{1}(m)+t_{2}(m),\cr
    b(m)&=&t_{1}(m)+b_{2}(m),\cr
    c(m)&=&b_{1}(m)+t_{2}(m)+t_{1}(m)+b_{2}(m).
    \label{abc}
\end{eqnarray}
The tunneling rates are 
\begin{eqnarray}
    b_{i} (n)&=&\left[\Delta E_{i}^{-}(n)/e^{2} R_{j}\right]/
\left[1-e^{-\Delta E_{i}^{-}(n)/k_{B}T}\right],\cr
t_{i}(n)&=&\left[\Delta E_{i}^{+}(n)/e^{2}R_{j}\right]/\left[1-e^{-\Delta 
E_{i}^{+}(n)/k_{B}T}\right],
\label{apprates}
\end{eqnarray}
where
\begin{eqnarray}
\Delta E_{1}^{\pm}(n)&=&\left[-{e}/{2}\ \pm\left(en-C_{j} 
V_{\rm ds} -C_{g} V_{g}\right)\right]{e}/{C_{\Sigma}},\cr
\Delta E_{2}^{\pm}(n)&=&\left[-{e}/{2}\ \mp\left(en+(C_{j}+C_{g}) 
V_{\rm ds} -C_{g} V_{g}\right)\right]{e}/
{C_{\Sigma}}.
\label{energies}
\end{eqnarray}
A convenient form for the tunneling rates is
\begin{eqnarray}
    b_{1}(m)&=&\frac{1}{R_{j}C_{\Sigma}}\frac{n-m+\Delta 
    n+\tilde{V}_{\rm ds}}{1-\exp\left[-(n-m+\Delta n+\tilde{V}_{\rm 
    ds})/\tilde{T}\right]},\cr
    b_{2}(m)&=&\frac{1}{R_{j}C_{\Sigma}}\frac{m-n-1-\Delta 
    n+\tilde{V}_{\rm ds}}{1-\exp\left[-(m-n-1-\Delta n+\tilde{V}_{\rm 
    ds})/\tilde{T}\right]},\cr
    t_{1}(m)&=&\frac{1}{R_{j}C_{\Sigma}}\frac{m-n-1-\Delta 
    n-\tilde{V}_{\rm ds}}{1-\exp\left[-(m-n-1-\Delta n-\tilde{V}_{\rm 
    ds})/\tilde{T}\right]},\cr
    t_{2}(m)&=&\frac{1}{R_{j}C_{\Sigma}}\frac{n-m+\Delta 
    n-\tilde{V}_{\rm ds}}{1-\exp\left[-(n-m+\Delta n-\tilde{V}_{\rm 
    ds})/\tilde{T}\right]},
    \label{appconvrates}
\end{eqnarray}
where 
\begin{eqnarray}
    \Delta n&=&\frac{C_{g}V_{g}}{e} -\frac{C_{g}V_{\rm ds}}{2e} -n -
    \frac{1}{2},\cr
     \tilde{V}_{\rm ds}&=&\frac{C_{\Sigma}V_{\rm ds}}{2e},\cr
      \tilde{T}&=&\frac{C_{\Sigma}k_{B}T}{e^{2}},
      \label{dimdef}
\end{eqnarray}
and $n$, $n+1$ are the most probable island electron numbers, 
determined by the condition
\[
n<\frac{C_{g}V_{g}}{e}-\frac{C_{g}V_{\rm ds}}{2e}<n+1,
\]
which is equivalent to $-0.5<\Delta n<0.5$.

Under the condition $k_{B}T\ll eV_{\rm ds}$ (equivalently 
$\tilde{T}\ll\tilde{V}_{\rm ds}$), we have that $t_{1}(n)$, 
$b_{2}(n)$, $b_{1}(n+1)$, $t_{2}(n+1)\approx 0$ for small drain-source 
voltage, and furthermore $b(m)\approx 0$ for all $m\leq n$ and 
$a(m)\approx 0$ for all $m\geq n+1$. Because the probabilities 
$\sigma(m)\approx 0$ for all $m\neq n$,~$n+1$, only the terms 
$m=n$ and $m=n+1$ need be considered in the summation over $m$ 
in eqs.~(\ref{appSI}) and (\ref{U}). From the form  of the 
matrix (\ref{Gamma}) and the above approximate conditions on the probabilities 
and tunneling rates, we then also have that only $m'=n$ and $m'=n+1$ 
need be considered in the sum over $m'$. The matrix elements 
$B_{n,n}$, $B_{n,n+1}$, $B_{n+1,n}$, and $B_{n+1,n+1}$ contain only 
the tunneling rates $a(n)$, $b(n)$, $c(n)$, and $c(n+1)$ and it is 
valid to just consider $2\times 2$ matrices for $\Gamma$ and $B$:
\begin{equation}
    B={\rm Re}\left(\begin{array}{cc}
i\omega-c(n) & -b(n)\\
-a(n) & i\omega -c(n+1)
\end{array}
\right)^{-1}.
\label{matrix}
\end{equation}
The probabilities $\sigma(n)$ and $\sigma(n+1)$ are given in Ref.\ 
\onlinecite{blencowe2} as
\begin{eqnarray}
    \sigma(n)&=&b(n+1)/[a(n)+b(n+1)],\cr
    \sigma(n+1)&=&a(n)/[a(n)+b(n+1)].
    \label{rhon}
\end{eqnarray}
Substituting  (\ref{matrix}) and (\ref{rhon}) into (\ref{appSI}) and 
(\ref{U}) and neglecting all small terms, we get  
\begin{equation}
S_{I}(\omega)=\frac{e^{2}a(n)b(n+1)}{a(n)+b(n+1)}+\frac{e^{2}[f(n)-g(n+1)]}
{a(n)+b(n+1)}\frac{a^{2}(n) g(n+1) -b^{2}(n+1) 
f(n)}{[a(n)+b(n+1)]^{2}+\omega^{2}}.
\label{appcurrentnoise}
\end{equation}

\section{Force noise due to the fluctuating island charge}

The electrostatic interaction between the fixed island  
and cantilever electrodes of the gate capacitor is
\begin{equation}
    F=A[C_{j}(V_{\rm ds}-2V_{g})-ne+Q_{p}]^{2},
    \label{appforce}
\end{equation}
where $Q_{p}$ is the background charge and 
\begin{equation}
    A=C_{g}(2 C_{j}-C_{g})/2dC_{\Sigma}^{3}.
    \label{A}
\end{equation}
Define
\begin{equation}
    \tilde{Q}=C_{j}(V_{\rm ds}-2V_{g})=-(2n+1+2\Delta n)e C_{j}/C_{g},
    \label{Q}
\end{equation}
where the second equality follows from (\ref{dimdef}). Neglecting 
$Q_{p}$ in (\ref{appforce}), the force noise is given as
\begin{equation}
S_{F}(\omega)=A^{2}\left\{4\tilde{Q}^{2}e^{2}S_{nn}(\omega)+e^{4}
S_{n^{2}n^{2}}(\omega)-2\tilde{Q}e^{3}\left[S_{n^{2}n}(\omega)+
S_{nn^{2}}(\omega)\right]\right\}.
\label{appSF}
\end{equation}
Similarly to expression (27) in Ref.\ \onlinecite{korotkov}, we have
\begin{equation}
    S_{n^{\alpha}n^{\beta}}(\omega)=4\sum_{m,m'} 
    m'^{\alpha}B_{m',m}m^{\beta}\sigma(m).
    \label{correlations}
\end{equation}
Using expressions (\ref{matrix}), (\ref{rhon}), and 
(\ref{correlations}) to obtain $S_{nn}(\omega)$, 
$S_{n^{2}n^{2}}(\omega)$, and $\left[S_{n^{2}n}(\omega)+
S_{nn^{2}}(\omega)\right]$, the final result for the force noise is
\begin{eqnarray}
   S_{F}(\omega)=&&\frac{4A^{2}e^{4}a(n)b(n+1)}{a(n)+b(n+1)}\cr
    &&\times \frac{(2C_{j}/C_{g})^{2} (2n+1+2\Delta n)^{2} 
    +(4C_{j}/C_{g})(2n+1)(2n+1+2\Delta 
    n)+(2n+1)^{2}}{[a(n)+b(n+1)]^{2}+\omega^{2}}.
    \label{appSF2}
\end{eqnarray}
Note that both $S_{I}(\omega)$ and $S_{F}(\omega)$ have a Lorenzian 
spectrum.

\section{Simple estimate for the displacement noise}

A simple, approximate formula for the tunneling current 
is\cite{beenakker}
\begin{equation}
    I_{\rm ds} =I_{\rm ds, max}/\cosh^{2}\left[e C_{g}(V_{g,{\rm 
    res}}-V_{g})/2.5k_{B}TC_{\Sigma}\right],
    \label{appcurrent}
\end{equation}
where $V_{g,{\rm res}}$ is the value of the gate voltage at which the 
tunneling current is maximal:
\begin{equation}
    V_{g,{\rm res}}=\frac{V_{\rm ds}}{2}+(n+\frac{1}{2})\frac{e}{C_{g}}.
  \label{vgres}
\end{equation}
Subsituting  (\ref{appcurrent}) and the shot noise limit formula  
$S_{I}=2eI_{\rm ds}$ into (\ref{SXI}), we obtain
\begin{equation}
    S_{X}^{I}=\frac{1}{2eI_{\rm ds, max}}\left[\frac{2.5 k_{B}T d 
    C_{\Sigma}}{C_{g}(V_{g}-V_{\rm ds}/2)}\right]^{2}\frac{\cosh^{4}
    \left[e C_{g}(V_{g,{\rm 
    res}}-V_{g})/2.5k_{B}TC_{\Sigma}\right]}{
    \sinh^{2}\left[e C_{g}(V_{g,{\rm 
    res}}-V_{g})/2.5k_{B}TC_{\Sigma}\right]}.
    \label{appSXI}
\end{equation}
Using expression (\ref{current1}) for the tunneling current with 
$\sigma(n)=\sigma(n+1)=0.5$ and
expressions (\ref{appconvrates}) for the tunneling rates with $\Delta 
n=0$  gives for 
the maximum tunneling current:
\begin{equation}
    I_{\rm ds, max}=V_{\rm ds}/4R_{j}.
\label{currentmax}
\end{equation}
To make a simpler estimate of the displacement noise, we might 
replace the dimensionless, parameter-independent function 
$\cosh^{4}[x]/\sinh^{2}[x]$  by a constant $K$, say its minimum value which is 4. However, this 
function poorly approximates the correct doublet shape  and it is 
found numerically that the $K\approx 40$ provides a good fit. 
Thus we obtain the final result for the displacement noise due 
to the shot noise in the tunneling current:
\begin{equation}
    S_{X}^{I}=K\frac{2R_{j}}{e V_{\rm ds}}\left[\frac{2.5 
    k_{B}TdC_{\Sigma}}{C_{g}(V_{g}-V_{\rm ds}/2)}\right]^{2}.
    \label{appapproxSXI}
\end{equation}

A simple approximation to the maximum force  noise is obtained by  
setting $\Delta n=0$ in (\ref{appSF2}) and neglecting small terms:
\begin{equation}
    S_{F}(0)=(A^{2}e^{5}R_{j}/V_{\rm 
    ds})\left[C_{\Sigma}(2n+1)/C_{g}\right]^{2}.
 \label{approxsf1}
\end{equation}
Substituting in the definition (\ref{A}) for $A$ and using 
$2n+1=C_{g}(2V_{g}-V_{\rm ds})/e$ at maximum tunneling current, we 
obtain:
\begin{equation}
    S_{X}^{F}=\frac{Q^2 e^{3}R_{j}}{k_{\rm eff}^{2}V_{\rm 
    ds}}\left[\frac{C_{g}(2C_{j}-C_{g})(V_{g}-V_{\rm ds}/2}
    {d C_{\Sigma}^{2}}\right]^{2},
    \label{appapproxSXF}
\end{equation}
at the resonant frequency $\omega_{0}$ of the cantilever, which is 
assumed to be much smaller than the tunneling rates in magnitude.

\begin{table}
\caption{The resonant frequency $f_{0}$ and mass $m$ of crystalline 
Si cantilevers with different lengths ($l$) and fixed width 
($w=0.6~\mu{\rm m}$) and thickness ($t=0.15~\mu{\rm m}$).}
\label{Table 1}
\begin{tabular}{ccccc}
$l$ in $\mu{\rm m}$ & $1.25$ & $1.5$ & $2$ & $4$\\
$f_{0}$ in MHz & $117.2$ & $81.36$ & $45.77$ & $11.44$\\
$m$ in pg & $0.262$ & $0.315$ & $0.419$ & $0.839$\\
\end{tabular}
\end{table}

\begin{figure}
    \vskip 1in
\caption{Scheme of the rf-SET displacement detector.}
\label{Fig.1}
\end{figure}
\begin{figure}
    \vskip 1in
\caption{Displacement noise due to the tunneling 
current shot noise as a function of gate voltage (upper, doublet curves). 
The estimate of the lower limit on the displacement noise vs. gate 
voltage is also shown (lower, dashed line). The noise analysis 
assumes a 
symmetric SET at $T=74~{\rm mK} (0.01e^{2}/C_{\Sigma})$ with 
junction capacitance $C_{j}= 100~{\rm aF}$, 
junction resistance $R_{j}= 50~{\rm k}\Omega$, and static gate 
capacitance $C_{g}= 50~{\rm aF}$ 
(the gate capacitor gap is $d=0.1~\mu{\rm m}$). The drain-source voltage is 
$V_{\rm ds}=0.38~{\rm mV} (0.6e/C_{\Sigma})$. The displacement noise is 
determined at 117~MHz-the resonant frequency of the cantilever.}
\label{Fig.2}
\end{figure}
\begin{figure}
    \vskip 1in
\caption{Displacement noise due to the force noise from the 
fluctuating island charge as a function of gate voltage (lower curves). 
The estimate of the upper limit on the displacement noise is also shown 
(upper, dashed line). The noise analysis assumes the same parameters as 
in Fig. 2.  }
\label{Fig.3}
\end{figure}
\begin{figure}
    \vskip 1in
\caption{The total displacement noise as a function of gate voltage 
(upper, doublet curves). For illustrative purposes, the sample voltage range 
is chosen such that the force and current displacement noise contributions 
are comparable. The tunneling current is given by the lower, 
dashed line.}
\label{Fig.4}
\end{figure}
\begin{figure}
    \vskip 1in
\caption{The total displacement noise as a function of gate voltage 
for the different cantilever lengths listed in Table I. The 
same parameter values as in Fig. 2 are assumed. The displacement noise 
at the resonant frequency of the cantilever is determined at the gate 
voltage for which the tunneling current is fixed at half its maximum 
value.}
\label{Fig.5}
\end{figure}
\begin{figure}
    \vskip 1in
\caption{Total displacement noise for different junction 
capacitances. Except for the junction capacitances, the same parameter 
values as in Fig. 2 are assumed and the displacement noise is 
determined at the tunneling current half-maximum.}
\label{Fig.6}
\end{figure}
\begin{figure}
    \vskip 1in
\caption{The displacement noise for different percentages of the 
maximum tunneling current. The same parameters as in Fig.~2 are used.}
\label{Fig.7}
\end{figure}
\vfill
\eject

\mbox{\epsfig{file=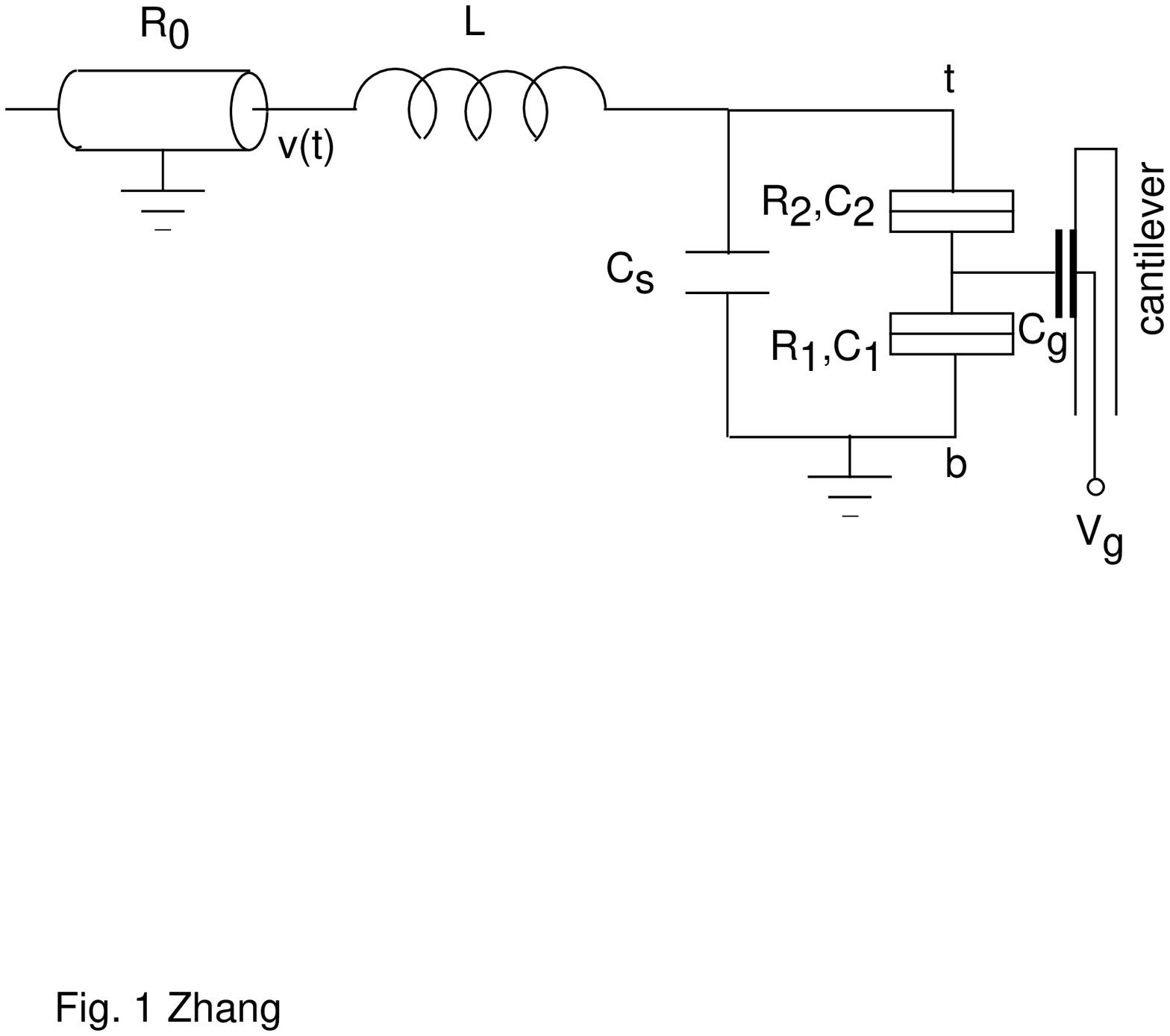, width=6in}}
\mbox{\epsfig{file=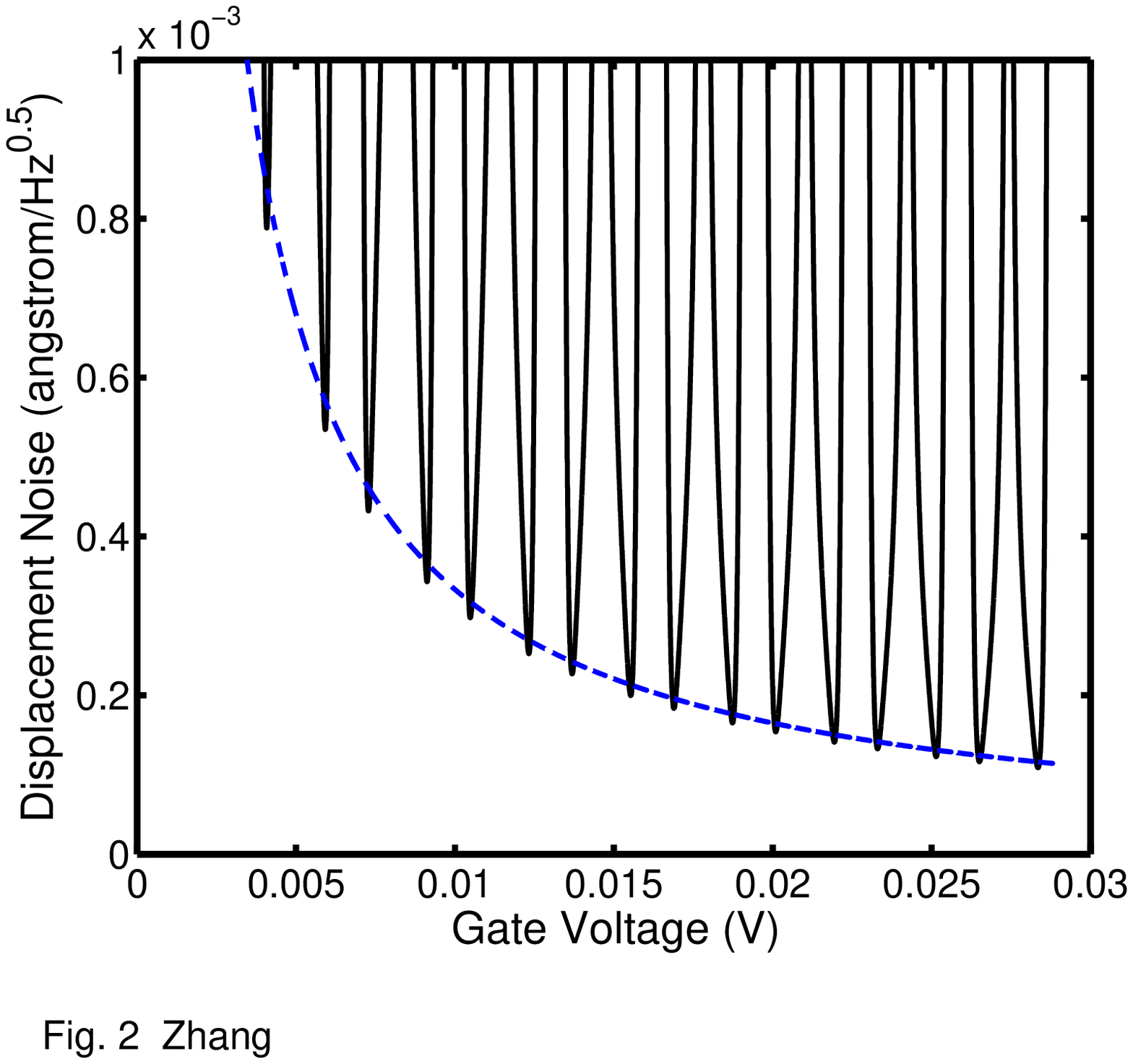, width=6in}}
\mbox{\epsfig{file=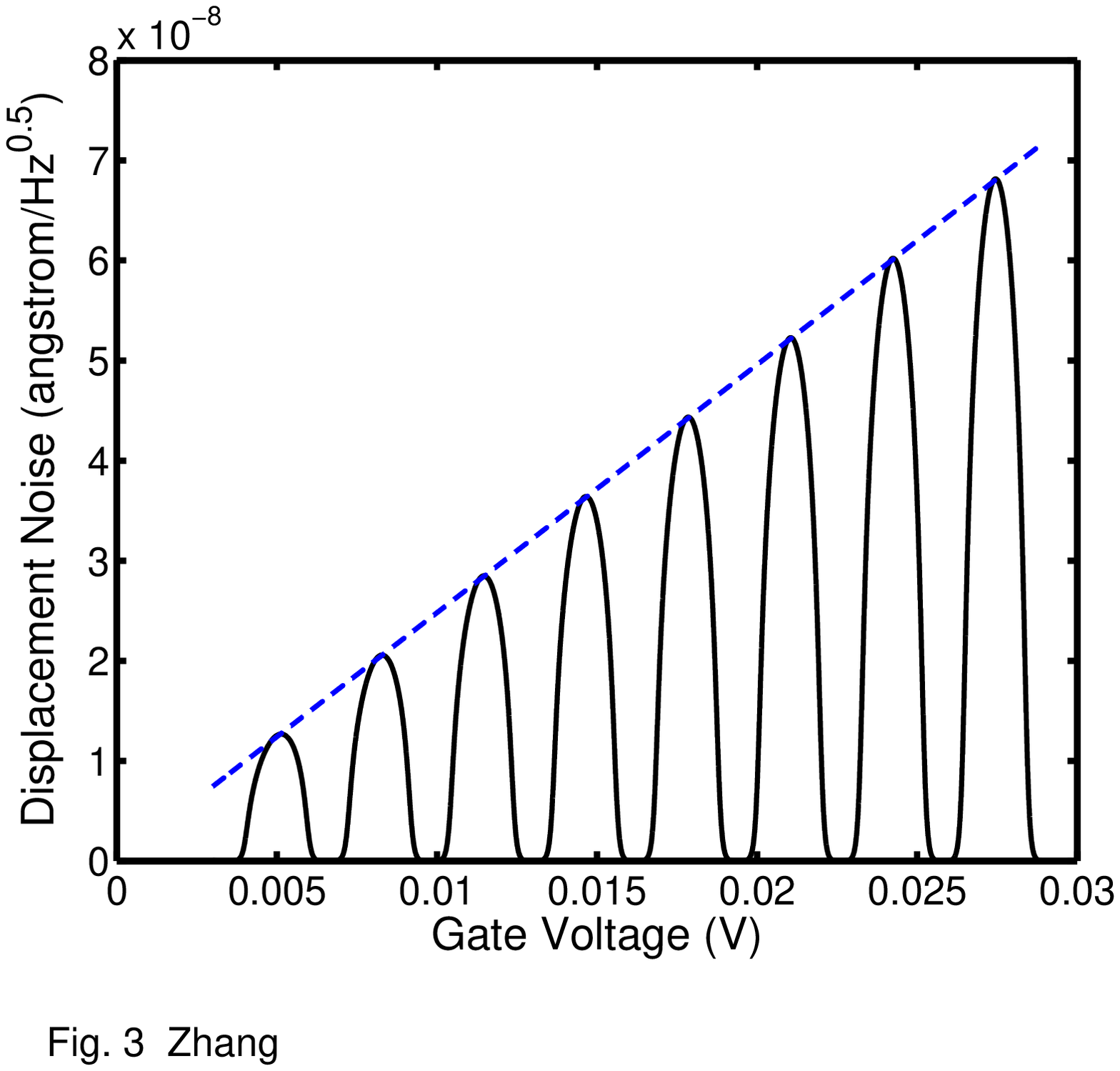, width=6in}}
\mbox{\epsfig{file=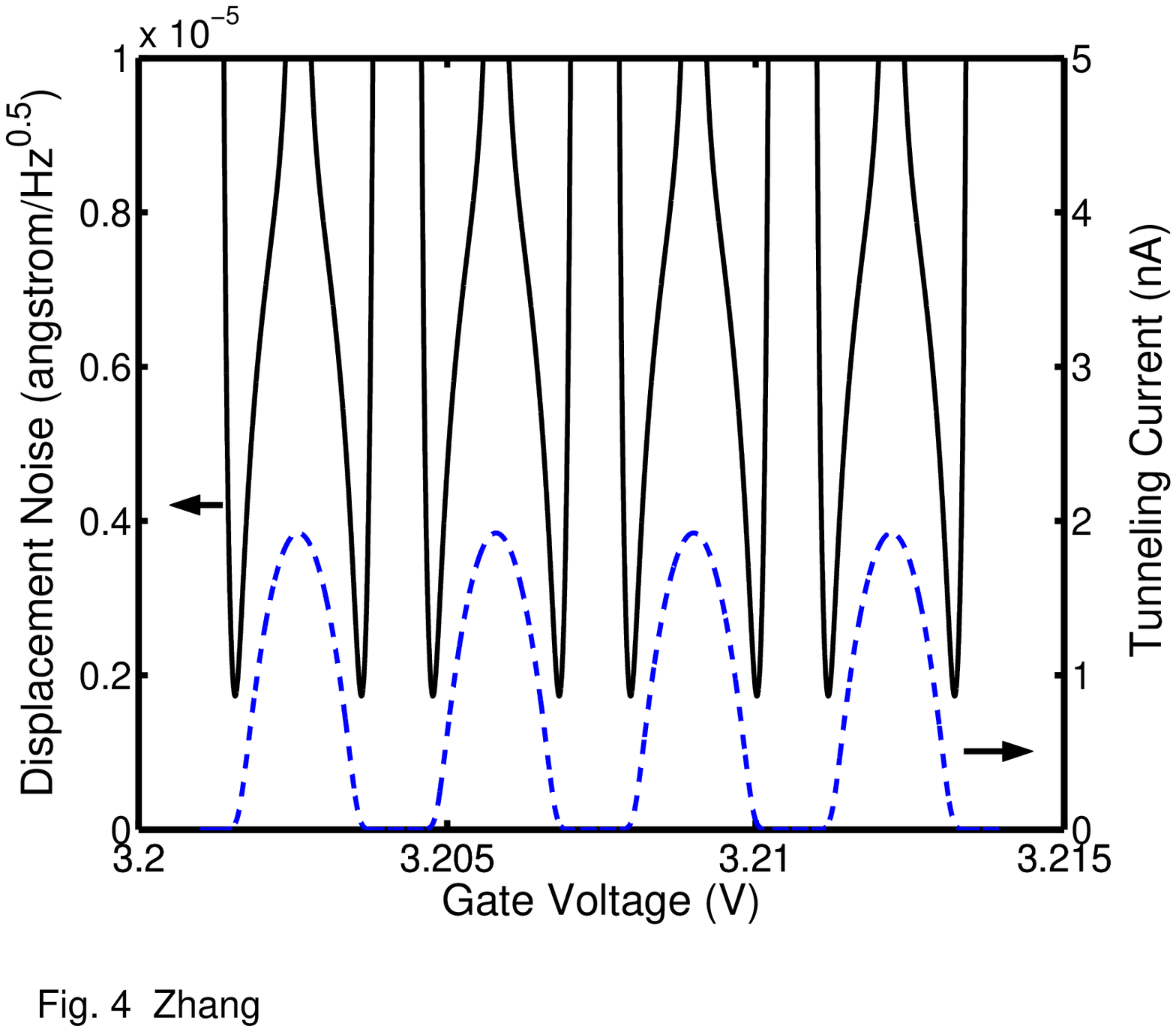, width=6in}}
\mbox{\epsfig{file=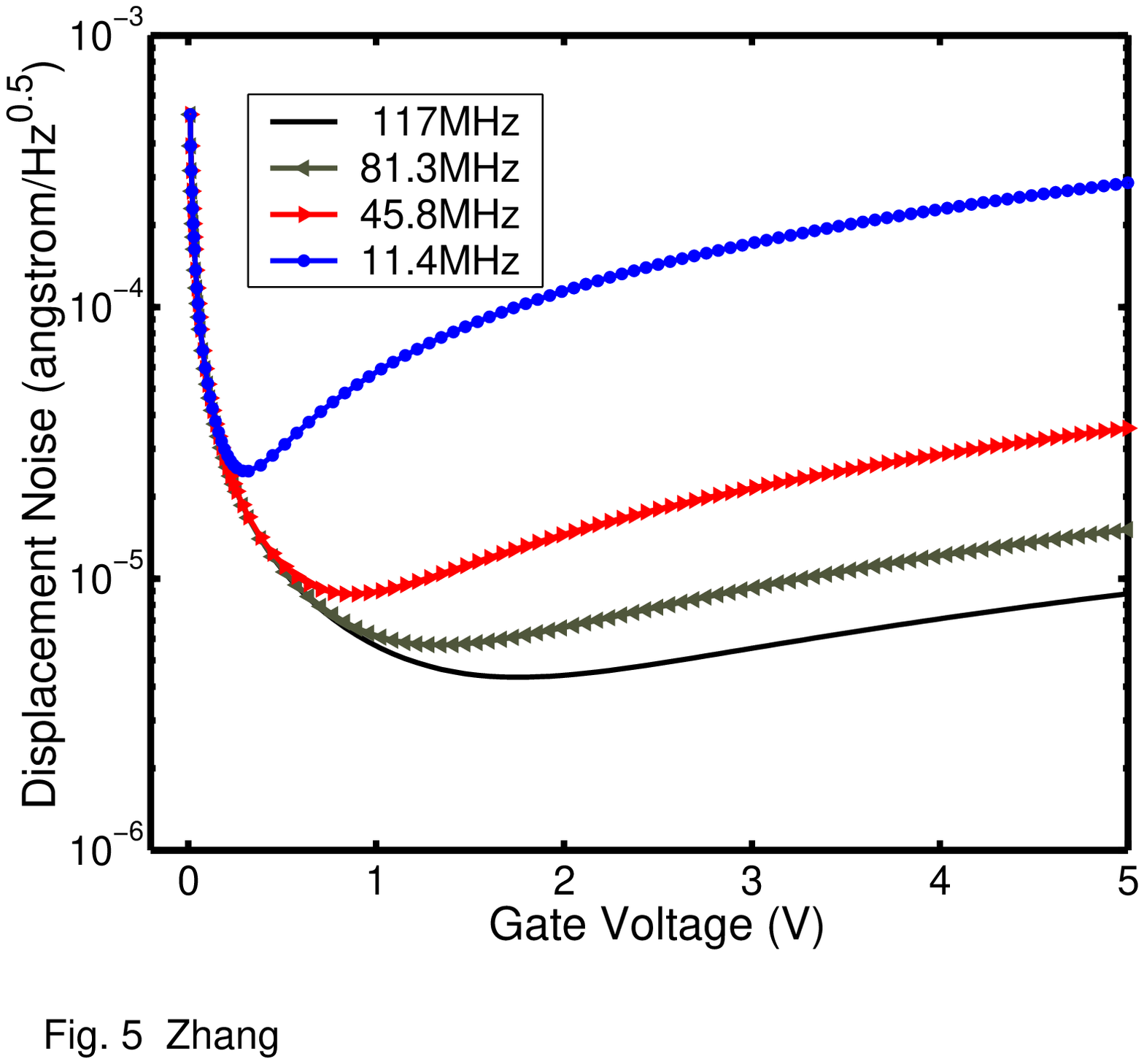, width=6in}}
\mbox{\epsfig{file=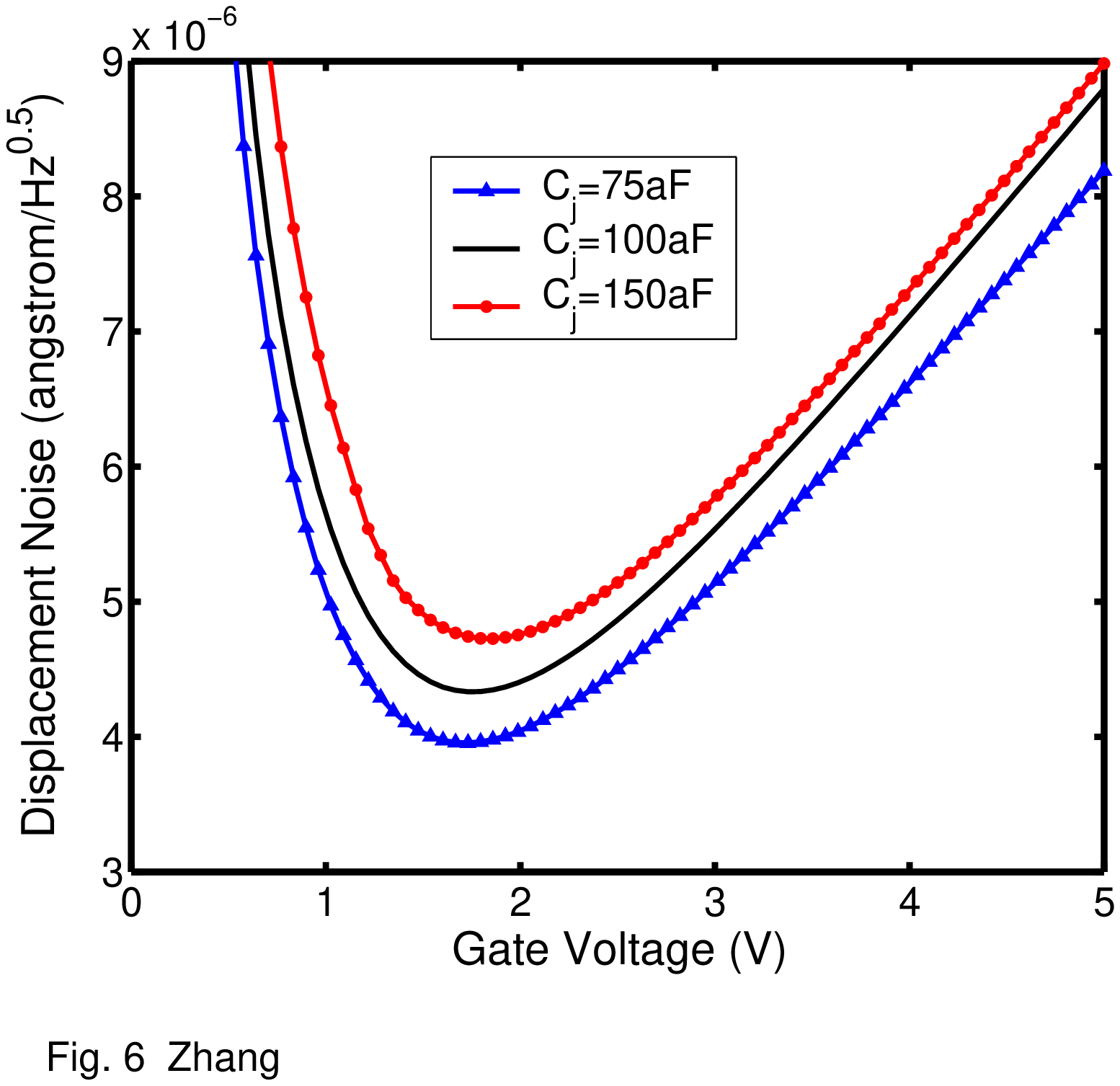, width=6in}}
\mbox{\epsfig{file=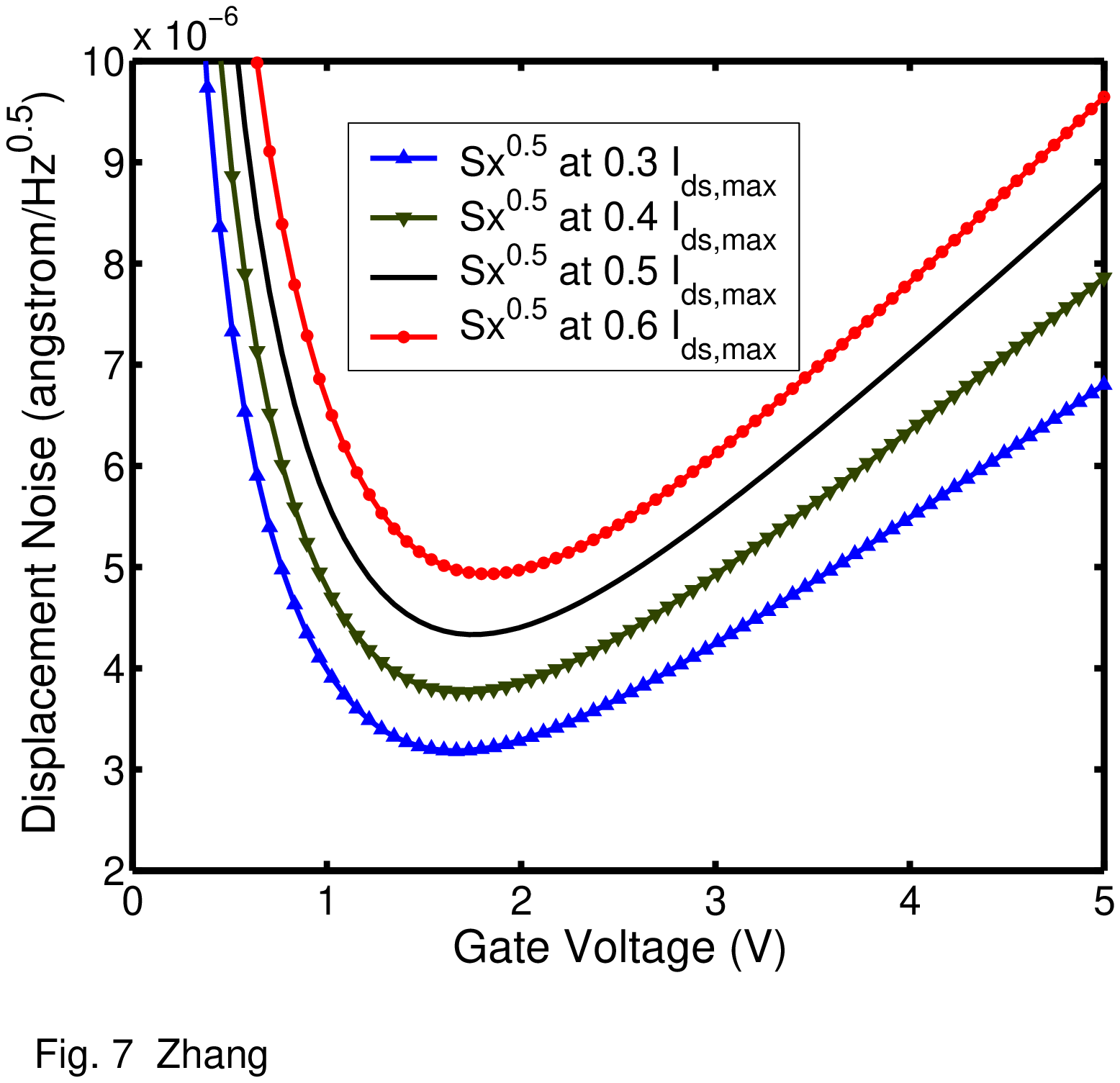, width=6in}}

\end{document}